\documentclass[conference]{IEEEtran}

\usepackage[T1]{fontenc}

\ifCLASSINFOpdf

\else

\fi

\usepackage{amsmath}

\usepackage[cmintegrals]{newtxmath}

\usepackage[noadjust]{cite}

\usepackage{caption}
\usepackage{algorithmic}
\usepackage{array}
\usepackage{float}
\usepackage{mdwmath}
\usepackage{mdwtab}
\usepackage{eqparbox}
\usepackage{ulem}
\usepackage{subfig}
\usepackage[font=footnotesize]{caption}
{\footnotesize
	\bibliographystyle{IEEEtran}}

\usepackage{lipsum}
\usepackage{kantlipsum}

\usepackage{makeidx}
\usepackage{amssymb}
\usepackage{multicol}
\usepackage{graphicx}
\usepackage{caption}
\usepackage{multirow}
\usepackage{framed}
\usepackage{enumerate}

\usepackage{color}

\newcommand{\subparagraph}{}

\usepackage[compact]{titlesec}
\titlespacing*{\section}{0pt}{2ex}{1ex}
\titlespacing*{\subsection}{0pt}{1ex}{0ex}
\titlespacing*{\subsubsection}{0pt}{0.5ex}{0ex}
\usepackage{titlesec}
\titleclass{\subsubsubsection}{straight}[\subsubsection]
\newcounter{subsubsubsection}
\renewcommand\thesubsubsubsection{\alph{subsubsubsection})}

\titleformat{\subsubsubsection}
{\normalfont\normalsize \itshape}{\thesubsubsubsection}{1em}{}
\titlespacing*{\subsubsubsection}{15pt}{0.25ex}{0ex}

\usepackage{url}
\makeatother

\begin{document}

%\title{Reinforcing SFC Security through Latency-Aware Virtualized Security Function Placement}
%\title{Introducing Latency-aware Virtual Security Functions for Securing SFCs}
\title{Introducing Virtual Security Functions into Latency-aware Placement for NFV Applications}

% author names and affiliations
% use a multiple column layout for up to three different
% affiliations
\author{
\IEEEauthorblockN{Ibrahim Tamim}
\IEEEauthorblockA{ECE Department\\
Western University\\
London ON, Canada\\
itamim@uwo.ca}
\and
\IEEEauthorblockN{Manar Jammal}
\IEEEauthorblockA{School of IT\\
York University\\
Toronto ON, Canada\\
mjammal@yorku.ca}
\and
\IEEEauthorblockN{Hassan Hawilo}
\IEEEauthorblockA{ECE Department\\
Western University\\
London ON, Canada\\
hhawilo@uwo.ca}
\and
\IEEEauthorblockN{Abdallah Shami}
\IEEEauthorblockA{ECE Department\\
Western University\\
London ON, Canada\\
abdallah.shami@uwo.ca}
}

%\vspace{-15em}}

%\vspace{-15em}
% make the title area
\maketitle

\begin{abstract}
		The shift towards a completely virtualized networking environment is triggered by the emergence of software defined networking and network function virtualization (NFV). Network service providers have unlocked immense capabilities by these technologies, which have enabled them to dynamically adapt to user needs by deploying their network services in real-time through generating Service Function Chain (SFCs). However, NFV still faces challenges that hinder its full potentials, including availability guarantees, network security, and other performance requirements. For this reason, the deployment of NFV applications remains critical as it should meet different service level agreements while insuring the security of the virtualized functions. In this paper, we tackle the challenge of securing these SFCs by introducing virtual security functions (VSFs) into the latency-aware deployment of NFV applications. This work insures the optimal placement of the SFC components including the security functions while considering the performance constraints and the VSFs' operational rules such as, functions' alliance, proximity, and anti-affinity. This paper develops a mixed integer linear programming model to optimally place all the requested SFCs while satisfying the above constraints and minimizing the latency of every SFC and the intercommunication delay between the SFC components. The simulations are evaluated against a greedy algorithm on the virtualized Evolved Packet Core use case and have shown promising results in maintaining the security rules while achieving minimum delays.\\
	\end{abstract}
	\IEEEpeerreviewmaketitle

	\section{Introduction} \label{Introduction}
\begin{figure} 
		\centering
		\hbox{\hspace{5ex}\includegraphics[width=3in, height= 2.1in]{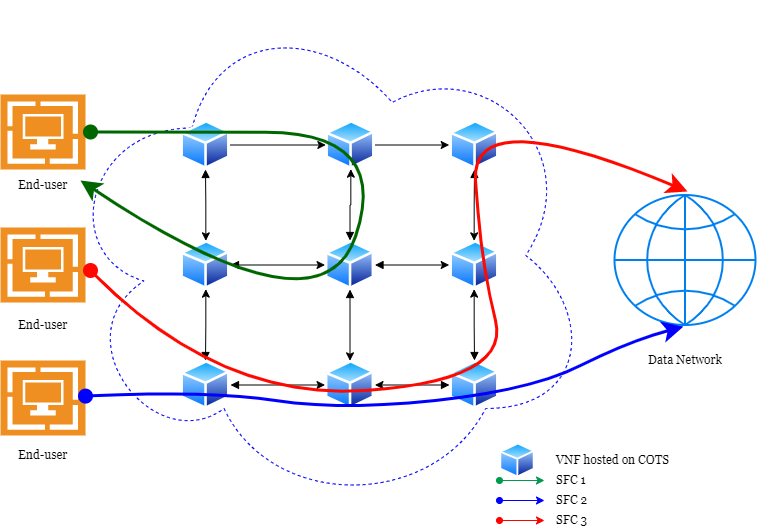}}
		\caption{SFCs in NFVI.}
		\label{SFCs}
		\vspace{-2em}
	\end{figure}

	With the exponential increase in networking demands, Network Service Providers (NSPs) are facing challenges such as, meeting strict delay requirements for mission critical applications, ensuring high-availability of network services, and securing vast amounts of data traversing the network. With their current legacy networking infrastructures, dealing with such challenges has proven to be very costly as it heavily increases the capital expenditures (CAPEX) and operating expenditures (OPEX) of the network provider \cite{mijumbi}. To mitigate these costs and tackle those challenges, NSPs are shifting to new models; Software Defined Networking (SDN) and Network Function Virtualization (NFV) \cite{jammal}. The objective is to use SDN and NFV as platforms for rapid expansion of their services, adding innovation, and lowering their costs while maintaining their Quality of Service (QoS) and tackling the aforementioned challenges. SDN and NFV have shown great potential in improving the economics of networking in parallel to giving the abilities to deploy, migrate, or terminate network services on-demand and optimizing the network performance by dynamically assigning resources when needed \cite{eramo}. \\
	%The key behind both concepts is virtualization where the network functionalities that are tightly coupled with their hardware infrastructures are transformed into software components that can run on any commercial-off-the-shelf (COTS) infrastructure. This transformation has enabled immense benefits to the telecommunication industry, allowing rapid scalability, flexibility, and adding new verticals to their platforms. \\
	In an NFV Infrastructure (NFVI), NSPs can quickly adapt to the end-user requirements by instantiating Virtual Network Functions (VNFs) in data centers (DCs) or network edges in a way to optimize their network resources and satisfy the carrier grade requirements from low latency constraints up to the high availability of network services \cite{hawilo1}. These VNFs are chained together to form an NFV application/service where data traverses specific VNFs following a specific order to create a Service Function Chain (SFC). Fig.~\ref{SFCs} shows a set of sample SFCs that can be generated for a certain NFV application that can run on commercial-off-the-shelf (COTS) servers. We can clearly see this architecture implemented today in content delivery networks, Vehicle-to-Infrastructure communications, home environment virtualization, and many more \cite{hawilo2}. However, with all those benefits, there are many challenges that should be addressed such as, security, computing performance, and carrier-grade service assurance \cite{hawilo2}. \\
	This work aims at enhancing the security of NFV services to mitigate challenges that arise due to the incorrect or inefficient placement of those VNFs. The process of placing VNFs in a network environment can be very complex. With a large set of candidate nodes that can host VNFs, the choice must be optimized to comply with the network requirements and constraints such as, network delay, availability of resources, security-aware placement needs, cost, available bandwidth, and other QoS requirements. There has been significant research towards optimizing the placement of the VNFs \cite{chang}\cite{bagaa}. However, there is still the need for extensive research regarding how these SFCs can be secured while still maintaining the previously mentioned constraints. In an NFVI, many security functions are virtualized and placed to ensure the security of the network. These functions include Deep Packet Inspection (DPI), Virtual Private Network (VPN) server, Firewall, Intrusion Detection System (IDS), Intrusion Prevention System (IPS), and many other services. In most cases, these functions are treated like any other VNF instance and are placed while overlooking their operation mechanism requirements. This practice creates critical vulnerabilities as these services require specific conditions to operate optimally or even correctly. Such conditions include their placement location with respect to the source or the destination, the type of traffic that can be passed through them, and the link delays to their dependent components. These conditions must be satisfied to ensure that the network is using the best security approaches to provide protection against attacks on its services. \\
In this work, we aim to tackle the challenges of virtualized security functions' (VSFs) placement within an NFVI with the goal of ensuring their optimal operations by considering their functionality and non-functionality-based constraints. This paper proposes a solution to achieve this objective while ensuring that all the requested network services are placed in a manner that guarantees the minimal latency achievable for that SFC. We propose a Mixed Integer Linear Programming (MILP) optimization model to minimize the SFC delay by encapsulating the security, performance, and other QoS constraints of the SFC. The optimization model captures the security constraints of the requested security services, then finds the optimal placements of all the requested VNFs including the VSFs that achieve minimal latency across the SFC, and ensures that all the security functions are placed in a manner to insure their proper operation. The main contributions of this paper are summarized as follows: 
\begin{enumerate}[i)]
\item Provide a set of security rules to place VSFs along with the SFC instances. 
\item Capture the nature of SFCs, which is translated using the dependencies/redundancies between the VNF instances.  
\item Propose an optimization model that chooses the optimal placement for VNFs/VSFs to minimize their latency. 
\item Consider functionality specific constraints of the chosen VSFs within the SFC to ensure the correct operation of all the security functions within the SFC.
\item Capture the interactions between the VSFs and VNFs of the SFC.  
\end{enumerate}
The rest of this paper is organized as follows. Section~\ref{Background} presents the related work for security-aware NFV placement. Section~\ref{Motivation} gives an overview of the security services' challenges/requirements in NFV environment. Section~\ref{Modeling} defines the problem architecture, the MILP model, and its constraints. Section~\ref{simulation} discusses the evaluation results. Finally, Section~\ref{conclusion} concludes the paper and defines the future work. 
	\section{Related Work} \label{Background}
	The placement of VNFs has received close attention from the research community. This is mainly due to how critical the location of these VNFs is on their applications' performance. In delay strict applications, the location of certain VNFs can play a big role in the overall performance of the SFC. \\
	Many approaches have looked at how to optimally place such services and reduce the overall delay experienced by the network. Luizelli et \textit{al.} \cite{luiz} formalize the problem of optimal placement of VNFs and the chaining of SFCs using an integer linear programming (ILP) model to achieve optimality in both reducing the delay while optimizing the allocated resources. Their heuristic approach generates a 25\% reduction in the delay of the network with a resource over-provisioning of 4\%. Bouet et \textit{al.} \cite{bouet} consider the placement of one type of a virtualized security function, which is the Deep Packet Inspection. They optimize the placement of this service while minimizing the cost of the operator and meeting the traffic and security targets. Their results show a clear relation between the network structure and the cost with the time performance. Although there are several approaches that aim at placing such services while considering some performance-aware constraints, they overlook the security-aware constraints of those placements in conjunction with other key requirements such as latency, availability, and other performance-aware requirements. Bhamare et \textit{al.} \cite{deval} provide an overview of the optimization problems in SFCs where they highlight the need for more research in the area of SFC security. Therefore, it is critical to design a model that focuses on the security aspect of SFCs especially in the era of Big Data, Internet of Things (IoT), and the proliferation of cloud applications. \\
This paper addresses one of the key challenges of NFV, which is the security-aware placement. As discussed earlier, most of the existing VNF placement optimization approaches discard the characteristics of the security services they place. With that being said, this paper understands the nature of the NFV services' requirements where each one of those services has strict constraints that should be met to function properly and not jeopardize the network as a whole. The details of such constraints and scenarios are discussed in the below section. However, capturing these security constraints into a deployment solution is not enough if the applications do not achieve the required performance. There rise the challenge of keeping the network latency at a minimum; hence, this paper aims to optimally place those services for the goal of maximizing the network security while maintaining the absolute minimum delay achievable. 
	\section{Problem Overview} \label{Motivation}
	NFV has provided a dynamic way of service instantiation in real-time, which seamlessly allows cloud-based applications to become optimized and more customizable for the end-user needs. With such abilities, the SFC becomes capable of applying modifications in real-time such as, instantiating, migrating, or dropping specific VNFs to provide the end-user with their requested service \cite{hawilo1}. Within this architecture, many security services can be requested specifically by the end-user or deployed automatically by the controller to ensure the overall security of the SFC. These security services have strict functionality constraints that should be satisfied for correct operation. However, these constraints are often overlooked in the literature work where most of the placement optimization techniques focus on more general constraints such as, delay, energy, and cost of the NFV applications \cite{deval}. Within those models, there is no separation between the type of VNFs that are being placed where all the VNFs (including the VSFs) are deployed and managed using the same set of constraints. This approach results in many unfeasible placements of the VSFs that put the overall security of the SFC at risk \cite{cheriet}. \\
	Shameli-Sendi et \textit{al.} \cite{cheriet} have proposed best practices for security functions by expressing them using Network Security Defense Patterns (NSDP). These NSDPs provide a general overview on how specific security services must be located in reference to each other or to other VNFs of an SFC. However, these constraints introduce potential delays on the SFC and sometimes contradict with the performance constraints of the services offered to the end-user. This can be very crucial for mission critical and delay sensitive applications. \\
	In this work, we propose a solution to address this problem by incorporating the security rules with the QoS constraints of the SFC to ensure that the SFC has the VSFs deployed using the best security practices while meeting the delay requirements. Our proposed solution considers four main security constraints that can be mapped to most VSFs in a cloud application. We namely consider alliance, conflict, redundancy, and proximity constraints. These constraints aim to ensure that all the VSFs of an SFC are placed in a manner that ensures their correct operation and does not compromise the network security. Our model takes into consideration the characteristics of each candidate service node (delay, available capacity, and interaction demands (affinity/anti-affinity zone)) with the goal of placing all the requested  VNFs and VSFs to achieve the lowest possible latency for that chain.
	\section{Problem Modelling} \label{Modeling}
	In order to incorporate security needs into the SFC placement, it is necessary to first understand the nature of the VSFs and their best practices.  This section describes the VSFs and their requirements and proposes the constraints to satisfy the QoS requirements and meet the service level agreement (SLA) in a cloud environment where a set of DCs in the same or in different zones are responsible of hosting the VNFs and VSFs.	
	\vspace{-1.1em}
	\begin{table} [!t] \label{threats}
		\footnotesize
		\begin{center}
		\vspace{0.8em}
			\captionsetup{justification=centering}
			\caption{Security threats and their counter security functions.}
			\label{threats}
			{\renewcommand*{\arraystretch}{1.3}
				\begin{tabular}{| >{\centering\arraybackslash}m{2in} |>{\centering\arraybackslash}m{1in} | }
					\hline
					\textbf{Security vulnerabilities} & \textbf{VSF} \\  \hline
					\begin{itemize}
					\item Excessive bandwidth usage \item Malformed network packets \end{itemize} & DPI \\  \hline
					\begin{itemize}
					\item Sybil attacks \item Denial of service attacks \item Policy violations \end{itemize} & IDS \\  \hline
					\begin{itemize}
					\item Data alteration \item	Sniffing attacks \item	Eavesdropping attacks \end{itemize} & Encryption \\  \hline
					\begin{itemize}
					\item Port scanning \item Fragmentation attacks \end{itemize} & Firewall \\  \hline
				\end{tabular}
			}
		\end{center}
		\vspace{-2em}
	\end{table}	
	\subsection{System design} \label{Architect}
	%\vspace{-1.1em}
	With the shift towards a virtualized environment, NSPs can mitigate many security threats. For instance, by using a centralized security management, the SDN controller can dynamically add, update, or terminate security policies across all VNFs. This is a major advantage in comparison to having per-VNF security policies that might become quickly outdated, as the architecture of the network and the neighboring VNFs is very dynamic. Therefore, it is important to capture all the VNFs instantiated within an SFC at once to determine the necessary security constraints required to manage the proper interaction between all the functions. Such an infrastructure is still susceptible to a wide range of security threats and attacks. Table~\ref{threats} shows a general overview of the most common security threats that NFV faces and the security function services that can be used to mitigate those threats.\\
In our work, we consider a set of key security functions to be placed within an SFC, namely Firewall, DPI, and IDS. It is necessary to note that these functions can also include encryption as a service, Network virus/malware scanning, Transport Layer Security (TLS) proxy, Distributed Denial of Service (DDoS) mitigation, data loss prevention, and application visibility \& control. A set of these services will be chosen for deployment within a requested SFC. The services can either be requested by the end-user explicitly or instantiated by the controller as default functions to secure certain network services. The overall placement of the entire SFC depends on the security and QoS constraints of the service nodes and on minimizing the latency of the SFC. Below are the details of each security constraint that is considered in our work.\\
\textit{a) Alliance:} This rule allows two VSFs to operate in parallel. It gives VSFs the ability to work together in terms of their functionalities to increase the security of the SFC.  \\
\textit{b) Conflict:} Through this rule, restrictions are added on which VSFs should not reside on the same node.\\
\textit{c) Redundancy:} For critical security services such as, IDSs and DPIs, it is important that they retain their states and information even after failure. Hence, this rule ensures that these services can recover through their redundant instances.\\
\textit{d) Proximity:} This rule insures that VSFs with strict delay requirements are deployed with the closest proximity to their dependent VNFs or VSFs by insuring the minimal delays on the links between them.
	\subsection{Mathematical modeling} \label{milp}
	Considering the above architecture, a MILP model is designed to ensure the optimal deployment of NFV services while maintaining security requirements and satisfying the performance needs in terms of latency, resources, and intercommunication constraints. This section gives an overview of the MILP model, its decision variables, objectives and the problem formulation. 
\subsubsection{Notations}
Each SFC has a set of VNF instances $v$ denoted as ${{I}^{v}}$. Different VSFs are placed in this chain to secure the VNF instances. Each set of security instances $s$ is denoted as ${{I}^{s}}$ . With that being said, a set of functions' instances is denoted as ${{I}^{T}}$ where $T$ represents the type of the instance whether it is a VSF or a VNF. The total number of instances in the ${{I}^{T}}$ set is defined as ${{I}^{T}_{total}}$.  ${{I}^{T}_{A}}$ represents a subset of ${{I}^{T}}$ where the instances here interact with each other in a distributed manner. In other words, this subset reflects the functions that satisfy the anti-affinity constraint (such as redundant instances). ${{I}^{T}_{co}}$ represents a subset of ${{I}^{T}}$ where the instances here closely collaborate with each other (dependency between instances). In short, this subset reflects the functions that satisfy the affinity/alliance constraint. Each server node belongs to a specific data center, and its set is denoted as $N$. ${{N}^{total}}$ represents the total number of servers in a given setting. A single node of the server set is defined as $n$. In terms of the computational resources representation, $C$ denotes the capacity set, which is CPU and memory. This set has two resource types, and each type is denoted as $c$.  A single instance of VNF or VSF is denoted as $i$. ${{I}^{Tcap}_{ic}}$ represents the resources requirement $c$ of a certain instance $i$ of specific type ${{I}^{T}}$. ${{N}^{cap}_{nc}}$ represents the available resources $c$ of a certain server node $n$. As for the delay parameters, the communication latency between two server nodes $n$ and $n'$ is defined as ${{L}^{Server}_{nn'}}$. ${{L}^{Tol}_{ij}}$ is the latency tolerance between the instances of ${{I}^{T}}$.
\subsubsection{Decision variables}
${{L}^{TT'}_{ij}}$ is an integer decision variable that defines the latency between two instances $i$ and $j$ of ${{I}^{T}}$ and ${{I}^{T'}}$ respectively where $T$ and $T'$ are of different types (either between VNFs, between VSFs, or between VNF and VSF). As for the placement binary decision variable, ${{D}^{T}_{in}}$ denotes the deployment state of a function instance $i$ on a server node $n$. The latency between the hosting nodes should meet the tolerance latency between functions' instances and is reflected using ${{Y}^{TT'}_{in}}$ . These decision variables are defined as follows:\\
\begin{equation}{
		{{D}^{T}_{in}}=\left\{ {\begin{aligned}
			& 1\ \ \ if\ n\ host\ i \\ 
			& 0\ \ otherwise \\ 
			\end{aligned}}\right.
	}\end{equation}
	\begin{equation}{
		{{Y}^{TT'}_{in}}=\left\{ {\begin{aligned}
			& 1\ \ \ if\ {{L}^{TT'}_{ij}}\ \leq\ {{L}^{Tol}_{ij}} \\ 
			& 0\ \ otherwise \\ 
			\end{aligned}}\right.
	}\end{equation}
\subsubsection{Optimization modeling}
The optimization MILP model aims to minimize the delay between the intercommunicating VNFs, between intercommunicating VSFs, and between VNFs and their corresponding VSFs. This objective function satisfies the aforementioned constraints, all for meeting the SLA and security requirements and the efficient operation/practices of the security instances while providing the optimal deployment of these functions. \\
\textit{Objective function:}
	%\vspace{0.6em}
	\begin{equation}{Minimize\quad \sum\limits_{i}^{{{I}^{T}_{total}}}\sum\limits_{j}^{{{I}^{T'}_{total}}}{{{L}^{TT'}_{ij}}}+\sum\limits_{i}^{{{I}^{T}_{total}}}\sum\limits_{j}^{{{I}^{T}_{total}}}{{{L}^{TT}_{ij}}}}
	\end{equation}
	\textit{Subject to:}\\
	%\vspace{0.6em}
	\textit{Decision Variables Constraints:}
	%\vspace{0.6em}
	\begin{equation}{
		{{D}^{T}_{in}} \& {{Y}^{TT'}_{in}} \ \epsilon \ \lbrace0,1\rbrace \quad \forall \ n\ \epsilon \,N,\quad \forall \,i\ \epsilon \ {I}^{T}
	}\end{equation}
	\begin{equation}{
		{{L}^{TT'}_{ij}} \geq 0 \quad \forall \ i,j\ \epsilon \,{I}^{T} \& {I}^{T'}
	}\end{equation}
	
	\hspace{-1em}\textit{Proximity Constraints:}
	%\vspace{0.6em}
	\begin{equation}
	\begin{aligned}({{L}^{Server}_{nn'}}-{{L}^{TT'}_{ij}}) \le M \times {{Y}^{TT'}_{in}} \\ 
	\forall \ i,j\ \epsilon \,{I}^{T} \& {I}^{T'} \ \forall \ n,n'\ \epsilon \,N \end{aligned}
	\end{equation}
	\begin{equation}
	\begin{aligned}({{D}^{T}_{in}}-{{D}^{T'}_{jn'}}) \le M \times (1-{{Y}^{TT'}_{in}}) \\ 
	\forall \ i,j\ \epsilon \,{I}^{T} \& {I}^{T'} \ \forall \ n,n'\ \epsilon \,N \end{aligned}
	\end{equation}
	\begin{equation}
	{{L}^{TT'}_{ij}}\ \le {{L}^{Tol}_{ij}}\quad \forall \ i,j\ \epsilon \,{I}^{T} \& {I}^{T'}
	\end{equation}
	\textit{where $T$ and $T'$ can be of different or similar function types}\\
	
	\hspace{-1em}\textit{Conflict \& Anti-Affinity Constraints:}
	\begin{equation}{
		{{D}^{T}_{in}}+{{D}^{T}_{i'n}} \leq 1 \quad \forall \ i,i'\ \epsilon \,{{I}^{T}_{A}},\quad \forall \,n\ \epsilon \ N
	}\end{equation}
	\begin{equation}{
		{{D}^{T}_{in}}+{{D}^{T'}_{i'n}} \leq 1 \quad \forall \ i,i'\ \epsilon \,{I}^{T} \& {I}^{T'},\quad \forall \,n\ \epsilon \ N
	}\end{equation}
	\textit{where $T$ and $T'$ are different function types} \\	
	
	\hspace{-1em}\textit{Collaborative Constraints:}
	\begin{equation}{
		{{D}^{T}_{in}}+{{D}^{T}_{i'n}} \leq 2 \quad \forall \ i,i'\ \epsilon \,{{I}^{T}_{co}},\quad \forall \,n\ \epsilon \ N
	}\end{equation}
	
	\hspace{-1em}\textit{Computational Capacity (Nodes) Constraints:}
	\begin{equation}{
		\sum\limits_{n=0}^{{N}^{total}}{{{D}^{T}_{in}}}=1\quad \forall \ i\ \epsilon \ {I}^{T}
	}\end{equation}
	\begin{equation}{
		\sum\limits_{i=0}^{{{I}^{T}_{total}}}{{{D}^{T}_{in}}}\,\times \,{{{I}^{Tcap}_{ic}}}\ \le \ {{N}^{cap}_{nc}}\quad \forall \,n\ \epsilon \,N,\quad \forall \,c\,\epsilon \,C
	}\end{equation}
\hspace{-0.2em}The proposed model generates optimal deployments of VNF instances and their corresponding VSFs where both functions will form the SFCs. This deployment takes into consideration the functionality constraints that are captured through the computational resources and latency requirements. It also considers the nature of security services and their interaction among each other (in terms of location, order, redundancy, and communication) as well as their relation with the SFC instances. This model also captures the dependency between the SFC components (VNFs and VSFs) as well as the intercommunication of the same VNF type instances. 
Constraints (4) and (5) ensure that the decision variables of deployments (${{D}^{T}_{in}}$ and ${{Y}^{TT'}_{in}}$) are binary and the latency is a positive one.
The latency constraints are captured in (6), (7), and (8). Constraint (6) ensures that the calculated delay should meet the latency tolerance between the SFC instances, the security services, or the VNFs and the security instances. Constraints (7) and (8) ensure that the SFC and security instances are deployed on server nodes that satisfy the latency requirements.
Due to the nature of VNFs and security services, there exist dependency and redundancy relations between them. Each of these interactions requires specific deployment settings. To ensure availability of at least one SFC and its corresponding security services, constraints (9) and (10) ensure that these redundant VNF instances or security instances should be distributed across different nodes. This can be within the same DC network or across different zones. These constraints can also apply on the VNF instances that collaborate closely and can tolerate the outage of the parent instance. Similarly, it applies on the security services that depend on each other with high tolerance time.  To enhance the QoS of SFCs and the communication among and within the security functions, constraint (11) ensures that the dependent VNF instances or security instances with low tolerance time (in case of outage/failure) can share the same server node. 
Regarding computational resource constraints, constraint (12) ensures that the server node has enough capacity to host the SFC's components. Constraint (13) determines that each instance whether a VNF or a VSF is deployed on at most one server node.
\begin{figure} 
		\centering
		\hbox{\hspace{1ex}\includegraphics[width=3in, height= 2.35in]{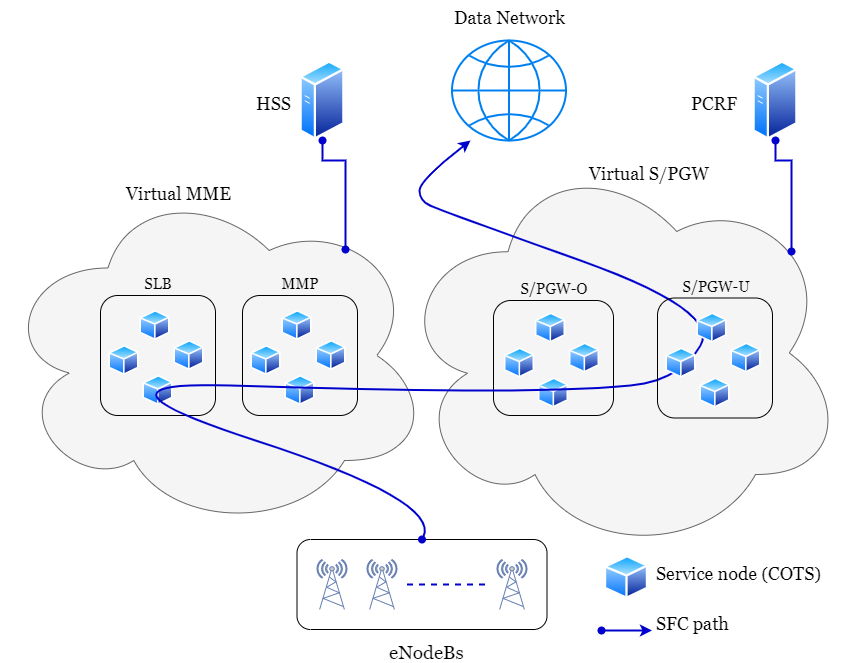}}
		\caption{vEPC Architecture.}
		\label{vepc}
		\vspace{-1em}
	\end{figure}
	\section{Use Case and Evaluation} \label{simulation}
	In order to understand the importance of incorporating delay requirements with security-aware constraints, we compare a latency-agnostic greedy algorithm with the proposed MILP model. The greedy algorithm aims at placing the VNF and VSF instances on the nodes that satisfy the computational resource requirements while overlooking the latency and the discussed security rules. This comparison highlights the importance of minimizing the SFC delays while insuring the correct and optimal placement of the requested security services. For the simulations, we have considered the virtualized Evolved Packet Core (vEPC) as a use case given its visionary nature that facilitates its extension to the emerging 5G system. Fig.~\ref{vepc} shows the vEPC architecture. The EPC is introduced as an all-Internet-Protocol (IP) core network architecture by the 3rd Generation Partnership Project (3GPP). It is used heavily in mobile networks for the purpose of providing broadband services \cite{etsi}. \\
	\begin{table} \label{dataset}
		\footnotesize
		\begin{center}
		\vspace{0.8em}
			\captionsetup{justification=centering}
			\caption{Simulation dataset.}
			\label{dataset}
			{\renewcommand*{\arraystretch}{1.5}
				\begin{tabular}{| >{\centering\arraybackslash}m{2in} |>{\centering\arraybackslash}m{1in} | }
					\hline
					\textbf{VNF/VSF Type} & \textbf{Number of Instances} \\  \hline
					MME & 2 \\  \hline
					HSS & 1 \\  \hline
					SGW & 2 \\  \hline
					PGW & 2 \\  \hline
					Firewall & 2 \\  \hline
					DPI & 2 \\  \hline
					IDS & 2 \\  \hline
					\textbf{Total} & \textbf{13} \\  \hline
				\end{tabular}
			}
		\end{center}
		\vspace{-2em}
	\end{table}	
\hspace{-0.4em}With SDN and NFV technologies, the major components of the EPC can be virtualized and decoupled from their hardware, allowing for rapid scalability and dynamic instantiation of those components in real-time to adapt immediately  to any end-user request or changes in network demands. Each main component is considered as a VNF, and the path from the end user through the vEPC components is the SFC. In this simulation, we consider the 4 network components of the vEPC, Home Subscriber Server (HSS), Mobile Management Unit (MME), Serving Gateway (SGW), and Packet Data Network Gateway (PGW). As for the security services, the VSFs to be deployed within the vEPC chain depend on the user requirements or can be pre-defined by the service provider. For this use case, we consider 3 VSFs, Firewall, DPI, and IDS, as they are essential security functions for the vEPC. For this combination of VNFs and VSFs, the model is given the set of intercommunication (dependency and/or redundancy) and security rules for each VNF and VSF. These rules are translated into the above constraints that should be satisfied to deploy the VNFs and VSFs. In this setup, the Firewall instances should be deployed with the closest proximity to their corresponding VNFs to ensure better network protection. The IDS instances should meet the alliance requirements where state sharing is enabled between IDS instances to protect certain VNFs and deal with variations in the attack traffic \cite{ids1}\cite{ids2}. In other words, the IDS instances are deployed as an active-active VSFs model on two different service nodes. Finally, the DPI instances are key functions that maintain critical information about the network traffic in their instances. Therefore and to ensure maximum protection of the SFC, these instances should be deployed with at least one redundant instance to recover their states in case of any failures. These redundant components should obey the anti-affinity rules. The structure of this simulation highlights the importance of considering the functional specific constraints for such critical VSFs while minimizing the delay across the generated SFCs and complying with the network requirements while enhancing its security. Table~\ref{dataset} shows the dataset (number of instances for each VNF and VSF) used to build the SFCs.
	\begin{figure*} 
		\centering
		\hbox{\hspace{3ex}\includegraphics[width=6.2in, height= 1.91in]{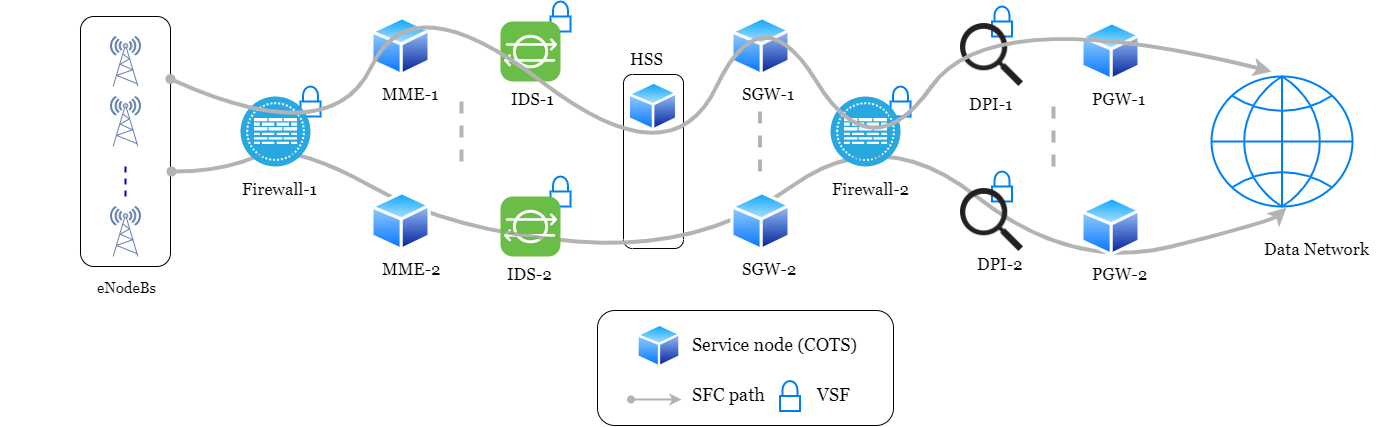}}
		\caption{The interaction of different instances (VNF \& VSF) of the SFCs.}
		\label{chainresult}
	\end{figure*}
	\subsection{Results and discussion} \label{results}
\begin{figure} 
		\centering
		\hbox{\hspace{-0.5ex}\includegraphics[width=3.5in, height= 2.2in]{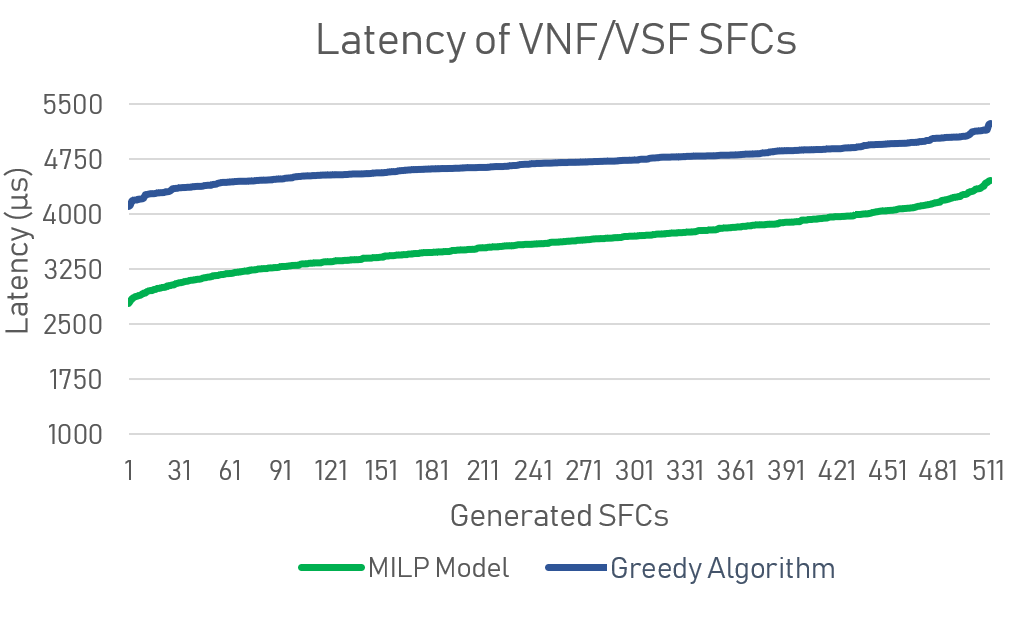}}
		\caption{The overall delay of the generated SFCs.}
		\label{overalldelay}
		\vspace{-1.4em}
	\end{figure}

\begin{figure} 
		\centering
		\hbox{\hspace{4ex}\includegraphics[width=3in, height= 1.63in]{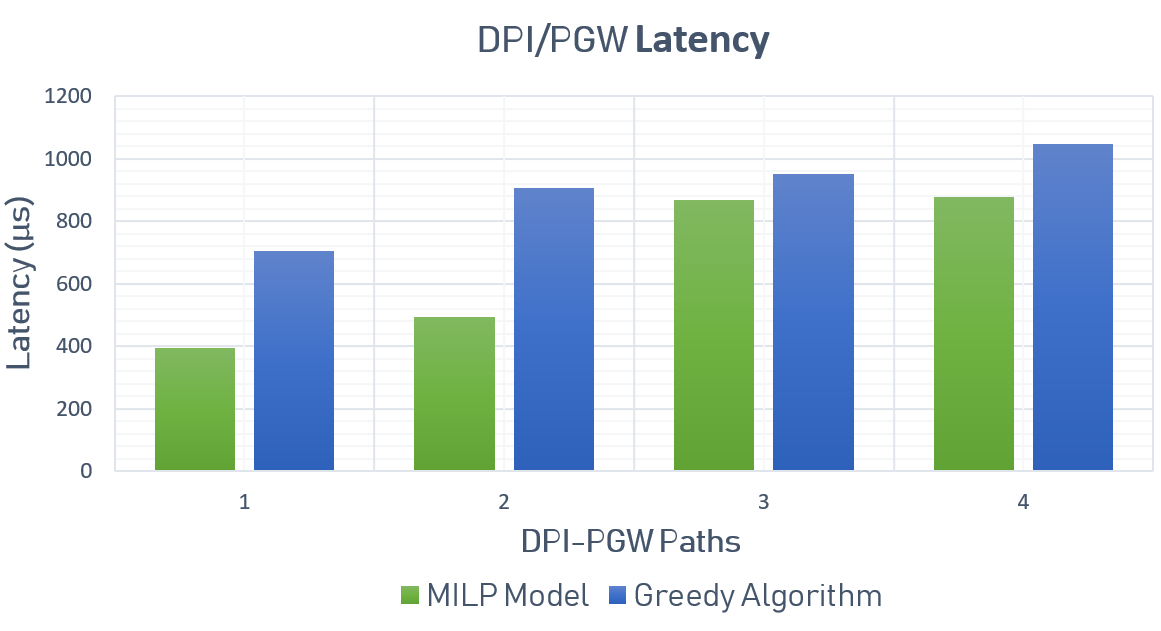}}
		\caption{The latency between DPI and PGW instances.}
		\label{intra1}
		\vspace{-1em}
	\end{figure}

\begin{figure} 
		\centering
		\hbox{\hspace{4ex}\includegraphics[width=3in, height= 1.57in]{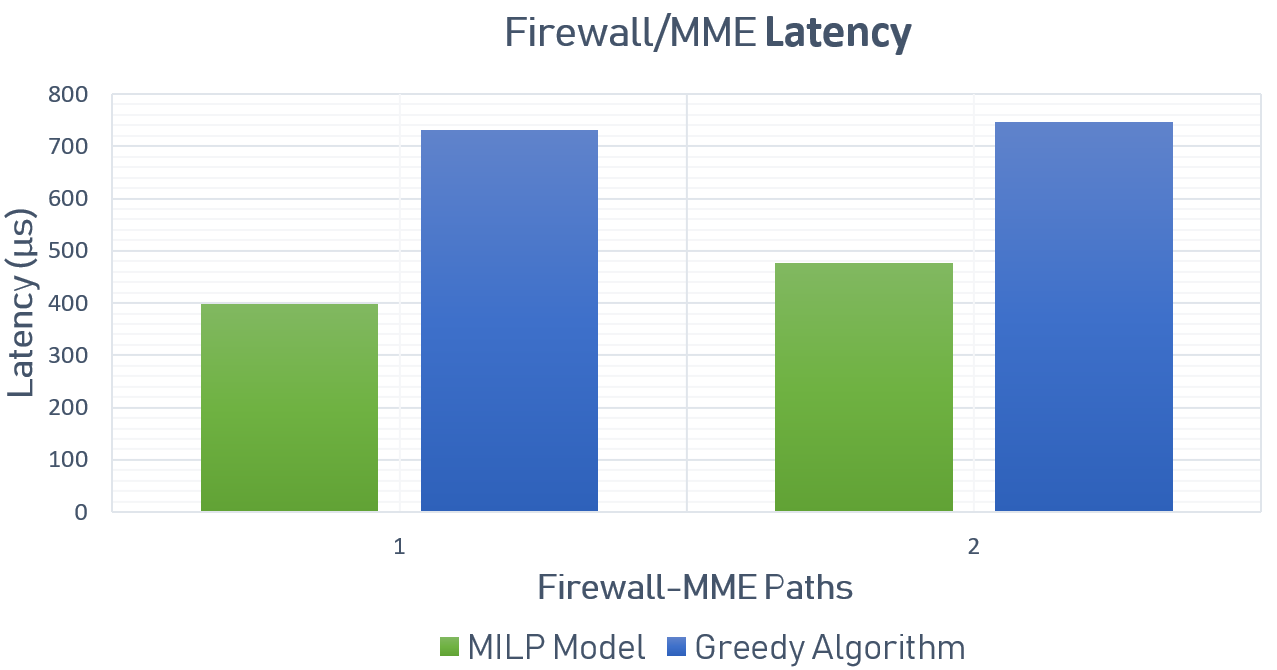}}
		\caption{The latency between Firewall and MME instances.}
		\label{intra2}
		%\vspace{-2em}
	\end{figure}

	The above simulation is evaluated through the vEPC core. It is carried on a Virtual Computing Server with 48 CPU cores and 128 GB of memory and has generated 512 SFCs. Fig.~\ref{chainresult} shows two of these optimal placement paths for each VNF and VSF. In this figure, the proximity rule has ensured that any instance of a Firewall is deployed with maximum proximity to its VNF. In this setting, the user nodes and the SGWs require Firewalls. The IDS instances are deployed as an active-active redundancy model and are insured to be hosted on different service nodes using the conflict/anti-affinity rule. Similarly, the DPI function has been deployed on a different service node than its redundant instance. The proposed model is capable of ensuring the minimal delay possible for all the 512 SFCs while traversing all their required VSFs. Compared to the latency-agnostic greedy algorithm, the proposed approach has achieved very promising results. Fig.~\ref{overalldelay} shows the latency in ($\mu s$) of all the generated SFCs and highlights how the MILP model has minimized their delays in comparison with the greedy algorithm.  The model does not only cover the minimization of the SFC latency as a whole, it also insures the minimal link delays between each VNF and VSF. These results give the controller and the user the flexibility of scaling on these links either by adding new VNF instances or enhancing the security of the SFC with introducing more VSFs. In terms of the intra-latency between the VNF and VSF, Fig.~\ref{intra1} shows the latency between the DPI and PGW instances using the proposed MILP model and the Greedy algorithm. Similarly, Fig.~\ref{intra2} highlights the link latency between the Firewall-1 and MME-1/MME-2. In the DPI-PGW case, we have 2 instances of each and thus 4 possible paths/links. In the Firewall-MME case, there is 1 instance of the firewall connected to 2 MME instances, which translates into 2 possible links. In both cases, the model significantly outperforms the greedy approach as it considers the latency of the interacting VNFs and VSFs as well as the whole chain. Table~\ref{intradelays} shows the detailed comparison of all the links between the VNFs and their corresponding VSFs. Our model is able to select the best deployments while satisfying the aforementioned security rules and minimizing the delay of every link created. With this evaluation, the proposed approach can be applied to other use cases in an NFV environment such as, Vehicle-to-Infrastructure communication services. It will also enable us to carry on with this work towards introducing additional security rules and enhancing the collaboration between VNFs and VSFs.
	\subsection{Computational complexity discussion} \label{complexity}
	Reduction methods are well-defined approaches that determine the complexity of optimization models. When a problem $Y$ is reduced to a certain problem $X$, the algorithm solving $X$ can be used for $Y$ as well. However, this method is not only used for design purposes, but also for classifying the difficulty or hardness of $Y$. Since the proposed problem falls in the scheduling category, it can be represented as bipartite matching graph problem. This graph has VNF and VSF instances as a set of vertices, the server nodes as the edges, and the deployment decision variable as the arc that maps a service instance to a node and denoted as $\lbrace i,n \rbrace$. Hoogeveen et \textit{al.} have reduced this graph problem to an NP-hard problem \cite{np1}\cite{np2}. With that being said, the proposed MILP model is an NP-hard problem by reduction. Thus, this problem can be solved by exact methods only for small-scale instances due to its time complexity. Since the exact solution cannot be determined by a polynomial-time approach for large scale, we aim at extending this work with an approximation, heuristic, or reduced enumeration algorithms to solve this challenge.
	\begin{table} [!t] \label{intradelays}
		\footnotesize
		\begin{center}
			\captionsetup{justification=centering}
			\caption{The latency ($\mu s$) between VNF and VSF instances.}
			\label{intradelays}
			{\renewcommand*{\arraystretch}{1.3}
				\begin{tabular}{| >{\centering\arraybackslash}m{1.5in} | >{\centering\arraybackslash}m{0.5in} | >{\centering\arraybackslash}m{0.5in} | }
					\hline
					Function Instances & MILP Model  & Greedy Algorithm  \\  \hline
					\textit{FW-1/MME-1} & 399 & 731  \\  \hline
					\textit{FW-1/MME-2} & 477 & 746  \\  \hline
					\textit{MME-1/IDS-1} & 417 & 721  \\  \hline
					\textit{MME-1/IDS-2} & 481 & 794  \\  \hline
					\textit{MME-2/IDS-1} & 640 &	959  \\  \hline
					\textit{MME-2/IDS-2} & 794 &	1047  \\  \hline
					\textit{IDS-1/HSS} & 417 &	959  \\  \hline
					\textit{IDS-2/HSS} & 794 &	1047  \\  \hline
					\textit{SGW-1/FW-2} & 715 &	823  \\  \hline
					\textit{SGW-2/FW-2} & 933 &	1083  \\  \hline
					\textit{DPI-1/PGW-1} & 394 &	706 \\  \hline
					\textit{DPI-1/PGW-2} & 492 &	905  \\  \hline
					\textit{DPI-2/PGW-1} & 869 &	951 \\  \hline
					\textit{DPI-2/PGW-2} & 877 &	1047  \\  \hline
				\end{tabular}
			}
		\end{center}
		\vspace{-2em}
	\end{table}	
	\section{Conclusion} \label{conclusion}
	SDN and NFV approaches have given NSPs massive incentives to quickly transform their architecture towards a fully virtualized environment. In such environment, NSPs should still comply with the strict SLA requirements and maintain their QoS while insuring maximum security of their networks. In this paper, we have addressed the challenge of securing the generated SFCs within a cloud environment by maximizing the effectiveness of the VSFs deployed to protect those SFCs. This paper has considered the strict functionality constraints of the VSFs to perform correctly. These constraints include alliance of functions, conflict, and proximity constraints. Additionally, this paper has also incorporated the strict delay requirements and other functional needs that should be met to comply with the SLAs. For this purpose, we proposed a MILP model to satisfy the security constraints of the SFCs while insuring the minimization of the SFC delays across the DC network. We compared the results with a Greedy latency-agnostic algorithm to pinpoint the importance of a benchmark model that includes the functional constraints and their impact on the NFV application's performance. Although the proposed MILP model has generated better results compared to the latency-agnostic approach, it is a computationally complex model. With this being said, this work will be extended to develop a heuristic solution for large scale network settings.

\end{document}